# Phase-sensitive imaging of microwave currents in superconductive circuits

A. Karpov,[1] A. P. Zhuravel,[2] A. S. Averkin,[1] V. I. Chichkov,[1] and A. V. Ustinov[1,3,4]

[1] *National University of Science and Technology MISIS, 119049 Moscow, Russia*

[2] *B. Verkin Institute for Low Temp. Physics and Engineering, NAS of Ukraine, 61103 Kharkov, Ukraine*

[3] *Physikalisches Institute, Karlsruhe Institute of Technology (KIT), 76131 Karlsruhe, Germany*

[4] *Russian Quantum Center, 100 Novaya Street, Skolkovo, Moscow region 143025, Russia*

The contemporary superconductive electronics is widely using planar circuits with micrometer-scale elements for a variety of applications. With the rise of complexity of a circuit and increased number of its components, a simple impedance measurement are often not efficient for diagnostics of problems, nor for clarifying the physics underlying the circuit response. The established Scanning Laser Microscope (LSM) technique generates the micrometer-scale images of the amplitude of the microwave currents in a planar superconductive circuit, but not the phase of the oscillating currents. Here we present a novel, more powerful type of LSM imaging containing the signal phase information. We employ a fast optical modulator in order to synchronize the phase of the laser intensity oscillation with the phase of the probing microwave signal. The loss induced in laser illuminated area strongly depends on the phase difference between the RF probing signal and the laser beam modulation. We explain the detection principle of the phase sensitive LSM and experimentally demonstrate the capability of this method using superconductive microwave resonators. The described technique facilitates understanding of complex RF current distributions in superconductive circuits.

The low-current superconductive electronics widely employs planar thin-film circuits for a variety of applications. To mention a few, these are devices for quantum metrology, quantum computing, ultra-sensitive THz detectors, SQUID magnetometers.[1-3] With the rise of complexity of circuits and increasing number of components, a simple impedance measurement may not be efficient for verifying circuit functionality. Employing spatially-resolved techniques for imaging of microwave currents makes it possible to better understand circuit operation regimes and its properties. In this respect, the Laser Scanning Microscopy (LSM) is a potent tool for imaging the RF current distributions in planar superconducting circuits with micrometer-scale resolution. The currently available LSM technique provides information on the local amplitude of the microwave currents in a circuit, but not the phase of these currents.[4-5] Making LSM imaging phase sensitive would significantly improve the understanding of the dynamics of interactions between different parts of the circuit and may allow for detecting complex modes of oscillations therein.

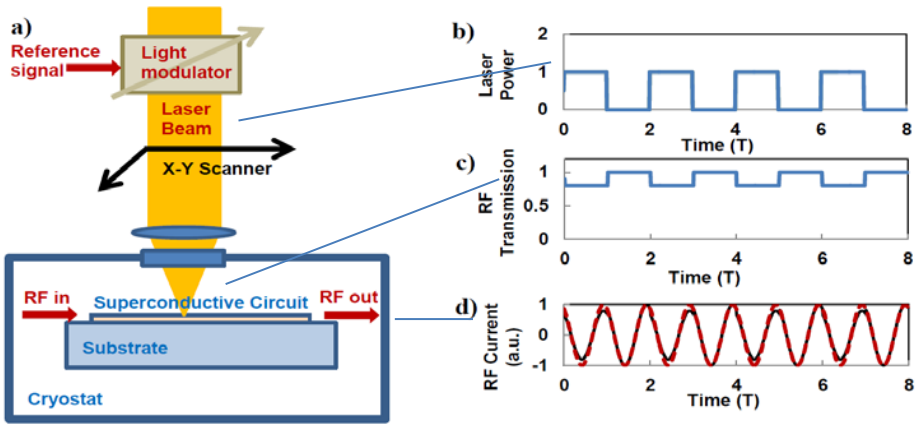

FIG. 1. The Laser Scanning Microscope is designed for visualizing RF currents in superconductive circuits. a) The schematics of LSM: A microwave signal at the frequency $f$ is transmitted through the superconductive planar circuit installed in a cryostat with an optical window. The laser beam is focused at the surface of the superconductive circuit at the position $X_1Y_1$ controlled by X-Y scanner. b) The laser beam intensity is pulse-modulated at the frequency $f/2$ phase locked with the RF source; c) The time profile of the loss induced by laser beam follows the beam intensity modulation, resulting in a small variation in the output RF signal. d) The dotted line is the input signal, and the continuous line is the output signal of the superconductive circuit. The phase of RF signal at $X_1Y_1$ location is shifted in respect of the laser modulation drive due to signal propagation in the superconductive circuit. As a result of parametric interaction in the illuminated area a signal is generated at the differential frequency ($f/2$) of the RF signal and laser modulation frequency. The amplitude and phase of the signal at $f/2$ are detected at the circuit RF output using a lock-in amplifier. Mapping of the amplitude and phase over the superconductive microcircuit is made with the help of an X-Y scanner.

The existing LSM technique employs a focused laser beam for scanning the surface of a superconducting circuit. In the area illuminated by the laser beam, the superconductivity is locally depressed by generation of the quasiparticles due to breaking of the Cooper pairs (photo response) and by heating effect of the laser radiation. The microwave currents flowing in the circuit are sensitive to quasiparticles due to the AC losses in the superconductor. The combined effect of laser radiation and microwave signal is the AC loss proportional to the square of the amplitude of the RF current.[4] In a typical LSM measurement, the transmitted (or reflected) microwave signal amplitude is registered and mapped as a function of the laser beam position, resulting effectively in imaging of the amplitudes of the microwave currents in the circuit.

The central idea of this work is to develop a novel, more powerful type of LSM measurement depicting the signal phase information. We are using a fast optical modulator for synchronizing the oscillations of the laser intensity with the frequency of the probing microwave signal. The loss induced in laser illuminated area of superconductive circuit is strongly dependent on the phase difference between the RF probing signal and the phase of the laser beam modulation.

The detection mode of microwave signal is essential for obtaining LSM response. In the traditional (scalar) LSM the transmission of the microwave signal is modulated at a relatively low frequency $f_M$ by pulse-driving the laser power supply. The amplitude of the induced pulsation of microwave signal is measured with a lock-in amplifier using a power meter at the circuit output port. Here, the reference oscillator driving the laser modulation is used as a reference for the lock-in amplifier.

Normally, the modulation frequency $f_M$ is chosen to be much smaller compared to the microwave frequency. The amplitude of pulsations of the output microwave power detected at the frequency $f_M$ constitutes the LSM response.

For the phase sensitive LSM mode described here, we use the same reference oscillator to phase lock the microwave signal source and the laser beam power modulation. As a simple example, let us consider LSM operation with the microwave signal in the superconductive circuit oscillating at the frequency $f$ with the phase $\varphi$, and the laser beam pulse-modulation at the frequency $f_M = f/2$ with the phase equal to zero (Fig. 1). Both the signal and the pulse modulation drive are phase locked to the same reference oscillator. While the $f/2$ frequency component is absent in the spectrum of the incoming microwave signal, it will be present at the circuit output port due to modulation of the circuit transmission by the laser beam power. The parametrically generated $f/2$ signal will thus be detected with a lock-in amplifier using the driving signal of the laser beam modulator as a reference. Normally, only a small part of the superconductive circuit is affected by the laser beam, and therefore the signal $a(t)$ at the circuit output port is only slightly dimmed by the applied laser power by a small factor $\alpha \ll 1$ corresponding to the scalar LSM response amplitude. The signal at the output port of the superconductive circuit may be written as:

$$a(t) = (1 - \alpha b(t)) \sin(2\pi f t + \varphi) \ , \tag{1}$$

where $b(t)$ is the unitary rectangular pulse train with the frequency $f_M = f/2$ (Fig. 1b) that can be represented as a sum

$$b(t) = \frac{4}{\pi} \left( \sin\left(\frac{2\pi f}{2} t\right) + \frac{1}{3}\sin\left(3\frac{2\pi f}{2} t\right) + \frac{1}{5}\sin\left(5\frac{2\pi f}{2} t\right) + \cdots \right) \ . \tag{2}$$

Substituting (2) into (1) we obtain:

$$a(t) = \left(1 - \alpha \frac{4}{\pi} \left( \sin\left(\frac{2\pi f}{2} t\right) + \frac{1}{3}\sin\left(3\frac{2\pi f}{2} t\right) + \frac{1}{5}\sin\left(5\frac{2\pi f}{2} t\right) + \cdots \right)\right) \sin(2\pi f t + \varphi) \ , \tag{3}$$

where the $f/2$ spectral component of the microwave signal $a(t)$ is

$$a_{f/2}(t) = \alpha \frac{4}{3\pi} \cos\left(\frac{2\pi f}{2} t + \varphi\right). \tag{4}$$

The amplitude of the $f/2$ component in the microwave spectrum at the circuit output is proportional to the induced loss factor $\alpha$, and its phase is equal to the phase of the RF signal at frequency $f$ in the area illuminated by the focused laser beam. When detected with a lock-in amplifier, the both amplitude and phase of the RF signal in superconductive circuit are measured. In other words, the proposed operation mode consists of mixing of the optical modulation signal with the microwave signal and, in general, is not limited to the particular choice of modulation frequency $f_M = f/2$.

For the experimental demonstration of a phase sensitive LSM operation, we use a superconductive resonator formed by planar Archimedean spiral with multiple inner modes shown in Fig. 2a. The resonator can be excited by the magnetic RF field perpendicular to its plane. We use the two current loops terminating the input and output coaxial lines as LSM signal ports, as described in[6, 7] (see also icon in Fig. 2b). The spectrum of inner modes of the Archimedean spiral is described by the harmonic rule[8]:

$$f_n = nl/c \tag{5}$$

where $l$ is the total length of the resonator spiral line, $c$ is the speed of light in the media, and $n$ is the mode number ($n = 1,2,3...$). It was also found that the radial profile of the inner modes follows a simple expression:

$$I_n(r/R_e) = I_{max} \sin(\pi n (r/R_e)^2) \tag{6}$$

with precision better than 5%.[8] Here $I_n$ is the amplitude of microwave current of the inner mode number $n = 1,2,3..$, $R_e$ is the external radius of the Archimedean spiral and $r$ is the radial distance from the center of the spiral. As it is usual for distributed resonator, the phase of the microwave current changes the sign at the nodes of the standing waves of the inner modes (Eq. 6).

We envisage demonstrating the phase sensitivity of the LSM by detecting the phase sign inversion in adjacent antinodes of the RF currents of the higher modes in spiral resonator. For this purpose, we fabricated a compact NbN spiral resonator shown in Fig. 2a. Here the Archimedean spiral has 75 turns with external radius of $R_e$=1.5 mm. The spiral line is 10 micrometer wide and about 10 nm thick. The LSM imaging of the current waveforms of the standing waves was performed at the resonance frequencies of the spiral. The measured spectrum of the inner modes of the spiral is presented in Fig. 2b. The spiral is inserted into microwave line between two weakly coupled magnetic coils, as shown in the inset of Fig. 2b. At the frequencies of spiral resonances, the loop to loop coupling peaks.[6,7] The resonance frequencies of the first three modes are harmonically related in proportion close to 1:2:3, as in Eq.5, and were exited at about 143 MHz, 309 MHz, and 452 MHz. In this compact resonator, the ratio of the wavelength at the frequency of the fundamental mode to spiral diameter is about 700.

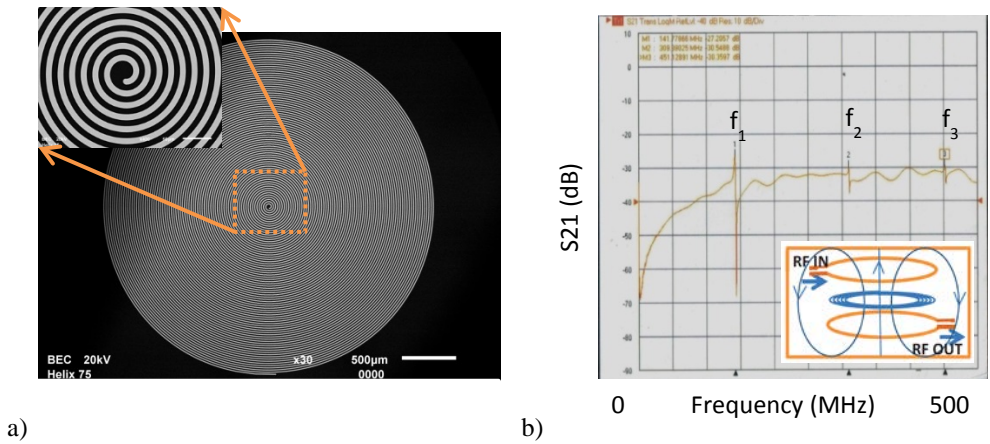

FIG. 2. Planar superconductive resonator shaped as Archimedean spiral. LSM measurements of the waveforms of the standing waves are performed at the resonance frequencies of the spiral. a) The spiral outer radius $R_e$ is 1.5 mm. The NbN spiral line is 10 micrometer wide and about 10 nm thick. b) The measured spectrum of the inner modes of the spiral. The spiral is connected to the microwave lines through two coupling loops, as shown in the insert. At the spiral resonance frequency the loop-to-loop magnetic coupling ($S_{21}$) peaks. The resonance frequencies of the first three modes scale approximately in proportion 1:2:3, as in Eq.5 (143 MHz, 309 MHz, and 452 MHz).

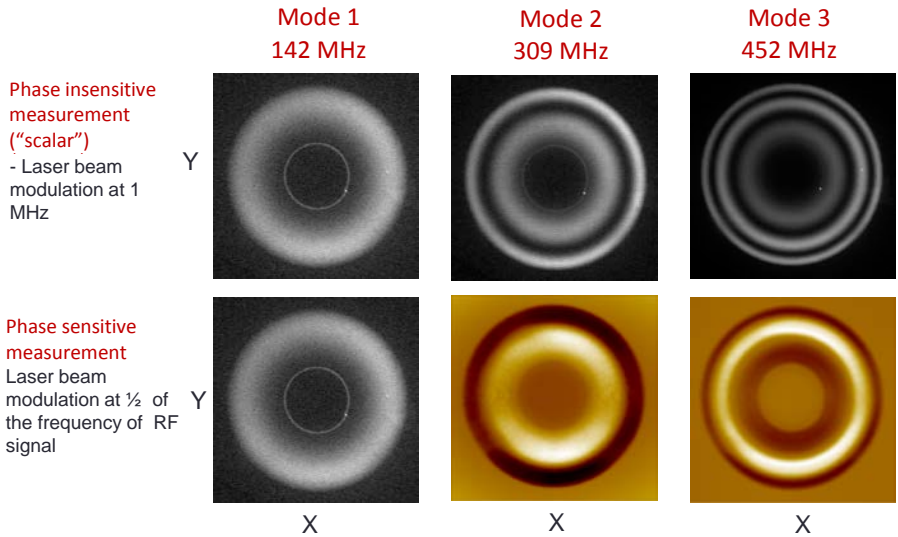

FIG. 3. Phase sensitive (bottom line) and traditional[4, 8] (top line) LSM imaging of XY distribution of RF currents at the 1-3 inner modes of a spiral resonator. The LSM optics spatial resolution is less than 20 micrometers, and the individual spiral turns are not resolved. The image coloring ranges from bright (higher LSM response amplitudes) to a color darker than the image background (negative LSM response). A bright spot indicate a short between the two turns of the spiral. In phase insensitive, traditional measurements the light beam is modulated at low frequency (1 MHz), and in phase sensitive regime the laser beam is pulse-modulated at ½ of the microwave signal frequency (*i.e.* at $f_n/2$). Notably, the images of the first mode current distribution are identical, while for the second and third mode they differ. The phase sensitive LSM images of the second and third modes demonstrate the contrast inversion in the adjacent antinodes.

The LSM images of the RF currents of the first three modes of the spiral resonator measured with the phase sensitive (bottom line) and conventional[4] (top line) LSM operation modes are presented in Fig. 3. The measured 2D distributions of RF currents at the three first modes in the spiral superconductive resonator are center-symmetric and present a combination of circular nodes and anti-nodes. The LSM spatial resolution here is about 20 μm, therefore the individual spiral turns are not resolved. A bright spot indicates a short between the two turns of the spiral. The coloring in the image ranges from bright (higher LSM response amplitudes) to a color darker than the image background (negative LSM response). The phase sensitive LSM data is at the bottom, the phase insensitive – in the top row of Fig. 3. In phase insensitive, conventional measurements the laser beam is modulated at low frequency (1 MHz), while in phase sensitive regime the beam is pulse-modulated at ½ of the microwave signal frequency (i.e. at $f_{n/2}$). Notably, the images of the first mode are identical, while at the second and third mode they differ. The latter two images demonstrate the phase inversion in the adjacent antinodes, confirming the expected operation in the phase sensitive mode.

The phase sensitive measurements with a higher optical resolution are illustrated in Fig. 4. With micrometer-scale LSM resolution we resolve the response of the individual turns of spiral line in resonator. Figure 4 presents radial line scans of the second and third modes in spiral resonator at frequencies 309 MHz and 451 MHz, respectively. The LSM response peaks at the crossings of the spiral turns and is close to zero between them. The phase sensitive LSM data are shown at the bottom (Fig. 4.c,d) and the conventional phase insensitive are presented in the top row (Fig. 4.a,b). As in the Fig. 3 data, for phase insensitive measurements the light beam is modulated at low frequency of 1 MHz, and in phase sensitive regime the laser beam is pulse-modulated at ½ of the microwave signal frequency. The envelope of the curves resembles to the model predictions given by Eq. 6. The phase inversion at the adjacent antinodes is clearly visible as alternation of the polarity of the response in Fig. 4c,d. The conventional scalar LSM signal shown in Fig. 4a,b produces a nearly identical profiles, but without change of the polarity.

The HFSS simulation of the current distribution at the 2nd and 3rd mode of the spiral resonator was used to verify the consistency of our interpretation of the LSM data. The measured and HFSS generated profiles of the signal currents along the radial direction are displayed in Fig. 4e,f. Here the HFSS data is drawn with blue line, the top curve at the edge of the spiral, where $r/R_e = 1$. For better perception the phase is inverted in HFSS simulated and measured profiles. The individual spiral turns are visible in both sets of data, giving a good reference for position. The short between the spiral turns in the central area is taken into account. It is visible as spike in HFSS curves located close to the center of the spiral, at $r/R_e = 0.2$. The measured amplitude of the RF currents is obtained as a square root of the LSM response. The curves envelopes, positions of the nodes calculated in HFSS match well to the observed in LSM data, and supports above interpretation the LSM results. For the 2nd mode, the node position is 14 turns away from the spiral edge, while for the 3rd mode the two nodes are located 7 and 26 turns away from the edge, respectively.

In summary, in this work we proposed and demonstrated experimentally a *phase sensitive* LSM technique for visualization of RF currents in superconductive microwave circuits. In order to obtain the phase contrast, we modulate the laser beam intensity at a frequency synchronized with a sub-harmonic of the microwave signal propagating through the circuit. The proposed technique for phase sensitive imaging of the microwave currents with micrometer-scale resolution opens way to explore and understand a great variety of complex RF current distributions and microwave patterns in superconductive circuits.

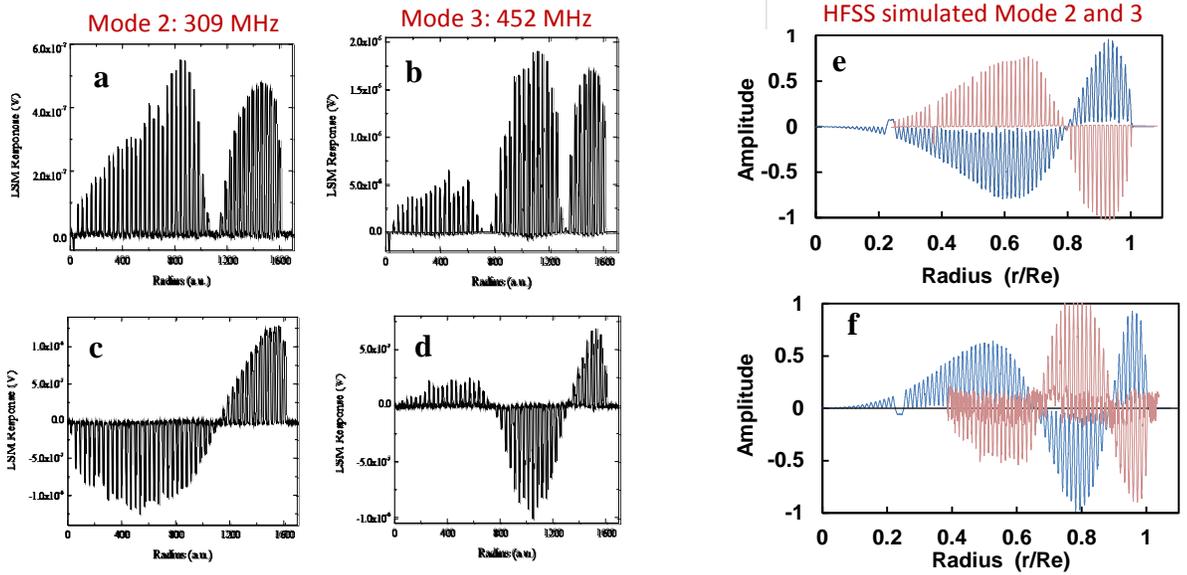

FIG. 4. A micrometer-scale resolution radial LSM scan from the center (around zero coordinate) to the edge of the spiral (around 1600 mark) in phase sensitive (bottom row: c, d) and insensitive (top row: a, b) technique. In the graphs are the measured profiles of LSM amplitudes of the standing waves of the second (a, c) and third (b, d) inner modes of a spiral resonator, both measured with phase sensitive (c, d) and scalar (a, b) LSM. The amplitudes are measured along a radius of the spiral and the individual spiral turns are resolved, making narrow spikes in LSM response. The envelope of the curve resembles to the model predictions of Eq. 6. The second and third mode resonance frequencies are 309 MHz and 451 MHz respectively. In scalar mode the LSM laser beam is modulated at 1 MHz, when in phase sensitive mode the laser beam modulation is applied at ½ of the signal frequency. The phase inversion at the adjacent antinodes is well visible when measured with the proposed phase sensitive version of the LSM. The scalar LSM measurements are giving a nearly identical profiles, but without phase inversion. e,f) The HFSS simulation of the current distribution at the second and third mode of the spiral (top curve at the edge of the spiral, when $r/R_e =1$) and the LSM measured data. Here the amplitude of RF current is obtained as a square root of the amplitude of the measured LSM response. The nodes' positions calculated in HFSS match well to the observed in LSM data (Mode 2: 14 turns from the edge and mode 3: 7 and 26 turns from the spiral edge). The short between the spiral turns in the central area is taken into account in HFSS model.

We would like to thank Steven M. Anlage for valuable discussions. We acknowledge the financial support of this work by the Volkswagen Foundation grant No. 90284, by grants No. 3.5490.2017/БЧ and 3.3360.2017/РН from the Ministry of Education and Science of Russian Federation, also grants No. K2-2014-025 and No. K2-2017-081 in the framework of Increase Competitiveness Program of the NUST MISIS.